# Technical Review

Imaging weak magnetic field patterns on the nanometer-scale and its application to 2D materials

Estefani Marchiori[1], Lorenzo Ceccarelli[1], Nicola Rossi[1], Luca Lorenzelli[2], Christian L. Degen[2], and Martino Poggio[1, 3, 4]

## Abstract

Nanometer-scale imaging of magnetization and current density is the key to deciphering the mechanisms behind a variety of new and poorly understood condensed matter phenomena. The recently discovered correlated states hosted in atomically layered materials such as twisted bilayer graphene or van der Waals heterostructures are noteworthy examples. Manifestations of these states range from superconductivity, to highly insulating states, to magnetism. Their fragility and susceptibility to spatial inhomogeneities limits their macroscopic manifestation and complicates conventional transport or magnetization measurements, which integrate over an entire sample. In contrast, techniques for imaging weak magnetic field patterns with high spatial resolution overcome inhomogeneity by measuring the local fields produced by magnetization and current density. Already, such imaging techniques have shown the vulnerability of correlated states in twisted bilayer graphene to twist-angle disorder and revealed the complex current flows in quantum Hall edge states. Here, we review the state-of-the-art techniques most amenable to the investigation of such systems, because they combine the highest magnetic field sensitivity with the highest spatial resolution and are minimally invasive: magnetic force microscopy, scanning superconducting quantum interference device microscopy, and scanning nitrogen-vacancy center microscopy. We compare the capabilities of these techniques, their required operating conditions, and assess their suitability to different types of source contrast, in particular magnetization and current density. Finally, we focus on the prospects for improving each technique and speculate on its potential impact, especially in the rapidly growing field of two-dimensional (2D) materials.

## Introduction

### Magnetic imaging

In the early 1800s, images of the stray magnetic fields around permanent magnets and current-carrying wires made with tiny iron filings played a crucial role in the development of the theory of electromagnetism. Today, magnetic imaging techniques continue to provide invaluable insights well beyond producing pretty pictures. They shed light on magnetization patterns, spin configurations, and current distributions, which are invisible in optical or topographic images. Unlike bulk measurements of transport, magnetization, susceptibility, or heat capacity, they provide microscopic information about length-scales, inhomogeneity, and interactions. Macroscopic manifestations of quantum mechanics involving strongly correlated states, e.g. superconductivity and magnetism, are sensitive to the local environment. In many cases, nanometer-scale spatial resolution is required to investigate and identify the conditions for their emergence.

---

[1] Department of Physics, University of Basel, 4056 Basel, Switzerland
[2] Department of Physics, ETH Zürich, 8093 Zürich, Switzerland
[3] Swiss Nanoscience Institute, University of Basel, 4056 Basel, Switzerland
[4] Email: martino.poggio@unibas.ch

Methods for magnetic imaging can roughly be divided into two categories: those sensitive to magnetization and those sensitive to stray magnetic field [1]. Magnetization mapping techniques include magneto-optic microscopy, synchrotron-based x-ray techniques, neutron diffraction, and electron polarization techniques. Spin-polarized scanning tunneling microscopy even allows the investigation of magnetic structure on the atomic scale. Direct measurements of magnetization are especially attractive for the investigation of magnetic domains and spin textures, because – in general – magnetization configurations cannot be uniquely deduced from stray field measurements alone. Nevertheless, for each of these techniques, a number of restrictions on the types of samples and conditions of measurement apply.

On the other hand, the generality of stray field measurements makes them applicable to a wider set of phenomena than magnetization measurements. Stray fields are produced not only by magnetization patterns, but also by current distributions. Furthermore, target samples span across conductors, insulators, and soft biological matter; the only requirement is that they produce a magnetic field. Common methods of mapping field include the use of fine magnetic powders as demonstrated by Bitter, Lorentz microscopy, electron holography, and a number of scanning probe microscopy (SPM) techniques. Foremost among these, for their high spatial resolution and high magnetic field sensitivity, are magnetic force microscopy (MFM), scanning superconducting quantum interference device (SQUID) microscopy, and scanning nitrogen-vacancy (NV) center microscopy.

By virtue of their outstanding properties, researchers are racing to apply these SPM techniques to investigate a range of emerging two-dimensional materials and van der Waals (vdW) heterostructures. Layer-by-layer stacking and twisting of these structures has recently opened a new area of materials engineering with huge potential for both fundamental discovery and applications. 2D materials can be metals, semimetals, topological insulators, semiconductors, insulators, and magnets. Correlation phenomena such as superconductivity, Mott insulating states, and magnetically ordered states have recently been observed in such materials and are just beginning to be understood. There is now an urgent need for sensitive and high-resolution SPM to zero-in on the nanometer-scale mechanisms behind these phenomena and to understand the role of disorder and lateral confinement in these new materials.

### Recent developments in 2D materials

The demonstration of the first graphene device in 2004 [2] launched the field of 2D and layered materials. Graphene itself, however, represents just one of manifold atomically thin 2D materials with a variety of compositions and crystal structures. Engineering these materials into heterostructures is made possible by the weak vdW interactions that typically dominate their interlayer coupling: these interactions allow the stacking and twisting of individual 2D layers without lattice mismatch adversely affecting the quality of the structure. This flexibility in both material choice and structure design has led to the synthesis and fabrication of 2D materials with properties ranging from insulating, semiconducting, metallic, to superconducting, and magnetic. The prospects for electronic, optical, and spintronic devices with new or superior functionality are great, especially because of the easy electrostatic control afforded by such thin structures.

The observation of long-range magnetic ordering in layered 2D vdW materials in 2017 [3,4] and the discovery of correlated insulator states and superconductivity in magic-angle twisted bilayer graphene (MATBG) [5,6] in 2018 are two particularly striking examples of the range and potential of 2D materials engineering. These phenomena are made possible because 2D vdW materials can be tuned into a flat-band regime where Coulomb interactions dominate over kinetic energy, putting the system in the strongly-correlated regime [7]. Engineering the structure can result in a system combining strong interactions and low electronic density, such that density can be controlled over the entire band using electrostatic gates. Such unparalleled control comes at the cost of sensitivity to disorder and inhomogeneity, which can result in the suppression of the desired correlated electronic state.

This sensitivity to disorder was made particularly clear by experiments correlating scanning SOT maps of twist-angle disorder with the electronic transport measurements in MATBG [8]. This result, as well as other scanning SOT experiments mapping quantum Hall edge currents in graphene [9,10] and imaging magnetization in twisted bilayer graphene [11], has demonstrated the importance of sensitive local magnetic field probes for understanding the underlying physics. Further work mapping layer-dependent magnetism in Cr-based vdW magnets using NV microscopy [12–14] has shown that more than one type of sensitive magnetic SPM may prove important for understanding 2D materials. Indeed, although MFM has not yet been applied to investigate correlated states or magnetism in 2D materials, techniques such as dissipation microscopy combined with new ultrasensitive NW MFM probes [15,16] are poised to make an impact.

### Outline

In this Technical Review, we describe the most promising state-of-the-art techniques for imaging current density and magnetization in 2D systems. We briefly discuss how each works and specify its magnetic sensitivity and spatial resolution limits. We also touch on the process of reconstructing spatial maps of measured magnetic field into images of current or magnetic moment. Finally, we compare the techniques and speculate on which is most suitable for which type of contrast and how each might best be applied in measurements of 2D systems.

## Background

### From iron filings to scanning SQUID-on-tip

One of the earliest techniques for visualizing magnetic fields involved dusting a sample with iron filings with their subsequent agglomeration revealing the form of the magnetic field lines. This method was refined by Bitter using magnetic nanoparticles in conjunction with either optical or electron microscopy to allow imaging of magnetic field lines with resolution down to 100 nm. Today, the highest spatial resolution is achieved by electron microscopy techniques, such as Lorentz microscopy or electron holography. In the former, high-energy electrons are transmitted through a thin sample; their deflection due to the Lorentz force depends on the magnetic induction within the sample. Similarly, using electron interferometry, electron holography maps the magnetic flux in and around magnetic samples. Although these techniques achieve spatial resolutions in the range of a few nanometers, their sensitivity to magnetic field is not high enough to resolve typical mesoscopic transport currents. Furthermore, Lorentz microscopy requires thin enough samples to be transmissive to high-energy electrons, typically on the order of 100 nm.

Near surfaces, the most common technique for imaging magnetic fields with high spatial resolution is MFM, which was introduced in the late 1980s as a natural extension of atomic force microscopy [17,18]. These days, it is performed in air, liquid, vacuum, and at a variety of temperatures. Because cantilevers are optimized to probe surfaces on the atomic-scale, they are designed to have spring constants around 1 N/m, which is smaller but on the order of spring constant of an atomic bond at the surface of a solid. As a result, conventional MFM can have extremely high spatial resolution, under ideal conditions down to 10 nm [19,20], but more typically from 30 to 100 nm. This large spring constant, however, makes MFM sensitive to strong magnetic field modulations on the order of tens of T/(m Hz$^{1/2}$) (few µT over 100 nm measured in 1 s). It is, therefore, well-suited for the measurement of highly magnetized samples, however, ineffective for detecting the weak stray fields produced by subtle magnetization patterns or Biot-Savart fields of currents flowing through nanometer-scale devices.

The advent of cantilever probes consisting of individual nanowires (NWs) [21,22] or even carbon nanotubes [23] have given researchers access to much smaller force transducers than ever before. This reduction in size implies both a better force sensitivity and potentially a finer spatial resolution [24]. Sensitivity to small forces provides the ability to detect weak magnetic fields and therefore to image subtle magnetic patterns; tiny concentrated magnetic tips have the potential to achieve nanometer-scale spatial resolution, while also reducing the invasiveness of the tip on the sample under investigation.

NWs have been demonstrated to maintain excellent force sensitivities around 1 aN/Hz$^{1/2}$ near sample surfaces (within 100 nm) when operated in high vacuum and at cryogenic temperatures, due to extremely low noncontact friction [25]. In recent proof-of-principle experiments, both magnet-tipped NWs and fully magnetic NWs were shown to be sensitive to magnetic field gradients of just a few mT/(m Hz$^{1/2}$) [10] and a few nT/Hz$^{1/2}$ [16], respectively. These are the gradients and fields produced by tens of $\mu_B$/Hz$^{1/2}$, where $\mu_B$ is a Bohr magneton, or several nA/Hz$^{1/2}$ of flowing current, each at a distance a hundred or so nanometers.

Other SPM techniques are typically applied when high sensitivity and low invasiveness are required. These include scanning Hall-bar microscopy, where efforts to develop non-perturbative probes have produced sensors less than 100 nm in size with a sensitivity of 50 µT/Hz$^{1/2}$ [26]. Micrometer-scale ultraclean graphene Hall sensors have been demonstrated with a sensitivity down to 80 nT/Hz$^{1/2}$ [27]. Scanning SQUIDs, however, are the scanning magnetic field probes with the highest sensitivity and bandwidth. Taking advantage of a SQUID's extreme sensitivity to magnetic flux, scanning SQUID microscopy was first realized in the early 1980s [28]. As imaging resolution has improved from the micrometer- down into the nanometer-scale, a number of strategies have been employed to realize ever-smaller sensors, which simultaneously retain high magnetic flux sensitivity and can be scanned in close proximity to a sample. One strategy has involved miniaturizing the pick-up loop of a conventional SQUID and placing it at the extreme corner of the chip where it can come close to a sample. The most advanced of such devices uses a loop with a 200-nm inner diameter to achieve sub-micrometer imaging resolution and a sensitivity of 130 nT/Hz$^{1/2}$ [29]. Although this design has the advantage of allowing for susceptibility measurements, the size of the sensor and minimum distance from the sample, which together determine the imaging resolution, are limited by the complex fabrication process. In the last decade, this limitation has been addressed through the development of SQUID-on-tip (SOT) sensors, consisting of a SQUID fabricated by shadow evaporation or directional sputtering of a metallic superconductor directly on the end of a pulled quartz tip [30,31]. This process has resulted in scanning SQUID sensors with diameters down to 50 nm, 100 nm imaging resolution, and a sensitivity of 5 nT/Hz$^{1/2}$ [32]. Scanning SQUID microscopy, which includes conventional and SOT sensors, has been used to image the stray field of two dimensional electron gases (2DEG) revealing coexistence of superconductivity and ferromagnetism [33], quantum spin Hall regime [34], and conducting edge currents [35]. Furthermore, the technique has unveiled quantized currents in mesoscopic rings [36,37] superconducting vortex dynamics [38,39], and equilibrium currents and magnetism in graphene [9–11].

As an important extension, SOTs have also been used for thermal sensing with a sensitivity below 1 µK/Hz$^{1/2}$. This sensitivity, due to the temperature-dependent critical current of the on-tip Josephson junctions (JJs), represents an improvement of four orders of magnitude over previous devices and allows for the detection and imaging of dissipation on the nanoscale [40]. Despite these advantages, the operating conditions required by these superconducting sensors limit their applications to cryogenic temperatures and high vacuum.

The last decade has also seen a flurry of activity in the development of scanning NV center microscopy. In this scheme, NV centers, which are optically addressable electronic defects in diamond, are used as scanning single-spin sensors. Following proposals in 2008 pointing out their potential for high-resolution, high-sensitivity magnetic field imaging [41,42], researchers have now demonstrated resolution better than 30 nm [43]. Using advanced sensing protocols and sequences of microwave and laser pulses, scanning NV center microscopes have achieved field sensitivities down to 4 µT/Hz$^{1/2}$ [44] for DC signals and 100 nT/ Hz$^{1/2}$ [45] for AC signals. Meanwhile, stationary NV centers reach 4 nT/ Hz$^{1/2}$ [46] and 1.3 nT/ Hz$^{1/2}$ [47] for DC and AC signals, respectively. NV microscopy has been used to image the stray field of magnetic vortices and domain walls [48–50], antiferromagnets and multiferroics [44,51–54], two-dimensional ferromagnets [12], helimagnets [55], skyrmions [56–58], superconducting vortices [59,60], nanometer-scale currents [43,61,62], and nanometer-scale ensembles of nuclear spins [63,64]. The best resolutions reported for scanning setups are between 15 and 25 nm [43,65], although resolution better than 10 nm should ultimately be possible for optimized scanning tips with very shallow NV centers. On top of high sensitivity and spatial resolution, scanning NV microscopy offers additional benefits: a large temperature range – including room temperature – a quantitative measurement of the magnetic field that is intrinsically calibrated via natural constants, vector sensitivity, and a number of spin manipulation protocols for performing spectroscopy from DC to GHz signal frequencies. Furthermore, NV microscopy can be extended to electric field [66–68] and temperature [69,70] measurements.

These advantages notwithstanding, scanning NV microscopy remains challenging at high fields due to the high microwave frequencies (10s to 100s of GHz) required to actuate the sensor electron spin, and the spin-level mixing for magnetic fields that are not aligned with the NV symmetry axis [71,72]. Although NV center detection has been reported below 1 K, experiments at cryogenic temperatures are hampered by reduced photoluminescence contrast and poor charge stability. Furthermore, the required optical excitation poses a limit on the possible samples, since it strongly perturbs materials such as direct-band-gap semiconductors, nanomagnets, and fragile biological structures.

## Imaging magnetization and current

Interest in nanometer-scale imaging of weak magnetic fields comes principally from a desire to non-invasively map charge transport and subtle magnetization patterns in nanostructures. Transport imaging is particularly useful for the direct observation of local disorder, bulk and edge effects, electron guiding and lensing, topological currents, viscous electron flow, microscopic Meissner currents, and the flow and pinning of superconducting vortices. Mapping of subtle magnetization patterns is important for investigations of antiferromagnetism, magnetic skyrmion phases, the spin-Hall effect, and magnetic phases in 2D materials.

Magnetic field probes detect currents and magnetization by virtue of the magnetic fields produced by moving charges and magnetic moments, respectively. Although, in general, a map of magnetic field cannot be reconstructed into a map of the source current or magnetization distribution, under certain boundary conditions the source can be uniquely determined. In particular, for 2D structures such as 2D materials, patterned circuits, thin films, or semiconductor electron and hole gases, a spatial map of a single magnetic field component can be used to fully reconstruct the source current or out-of-plane magnetization distribution. Since some of the most interesting and elusive effects are observed over length-scales of less than 1 μm and with currents less than 1 μA or magnetizations of few $\mu_B/nm^2$, techniques are required with both nanometer-scale spatial resolution and a sensitivity to fields smaller than a μT.

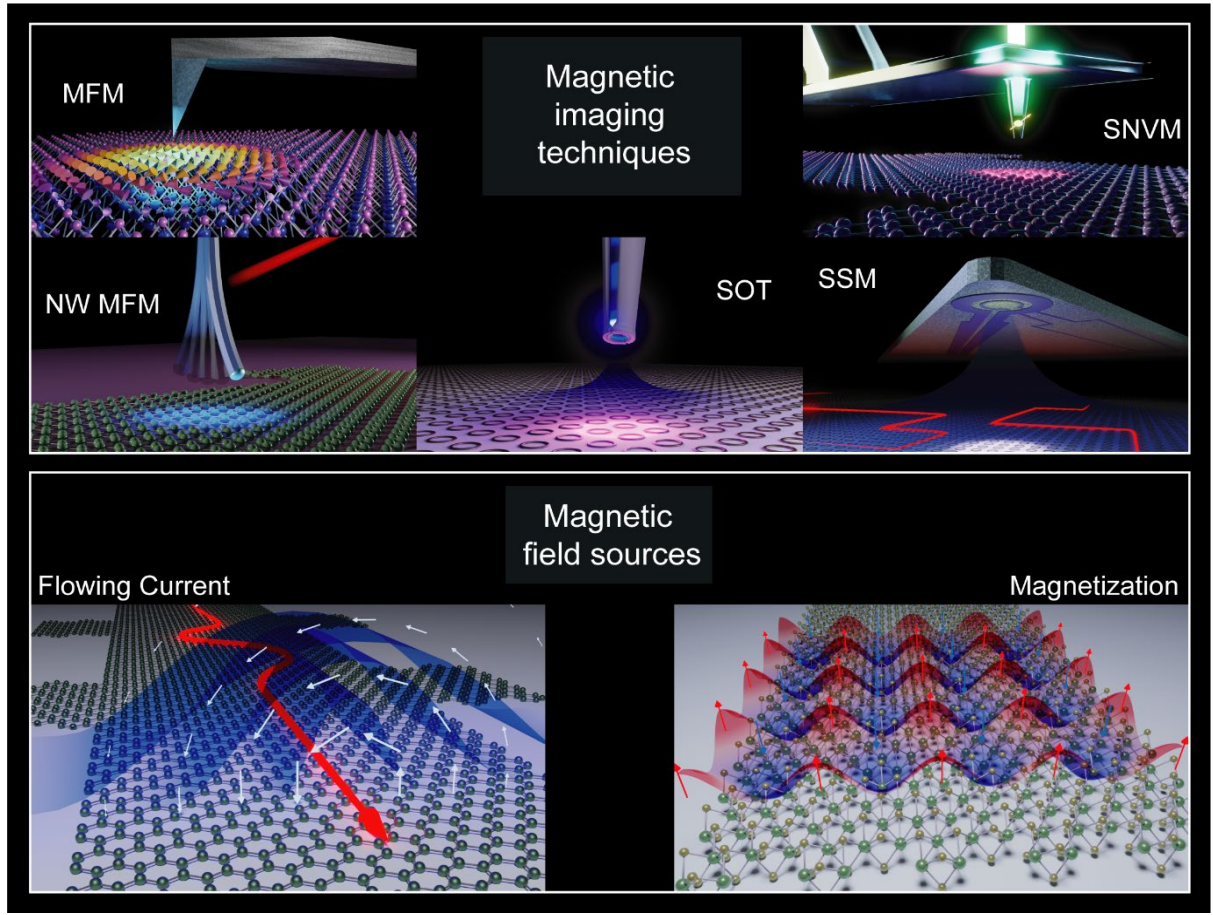

Figure 1: Schematic showing the principal magnetic imaging techniques and sources of magnetic field discussed in this review.

## Imaging weak magnetic field patterns with high spatial resolution

In SPM, high spatial resolution is achieved by minimizing both sensor size and its distance from the sample. High sensitivity is obtained by maximizing signal-to-noise ratio for the magnetic signal of interest and the fundamental noise of the measurement. In evaluating the sensitivity of different techniques to magnetic contrast, we follow Kirtley [73] and consider their response to two idealized sources of magnetic field: a magnetic dipole moment and a line of current. This procedure allows us to assess and compare the sensitivity of each technique to magnetization and current density in a sample below.

### Magnetic force microscopy
*Working principle and conditions*
In MFM, contrast results from the magnetostatic interaction between the stray magnetic fields of a sample and the magnetic tip of a mechanically compliant scanning probe. The vibration frequency and amplitude of a cantilever probe, whose tip has been coated with a ferromagnetic film, are recorded as the probe is scanned above a sample. The response typically depends on a gradient of the stray field. Although some simplifying assumptions can often be made, extracting exact magnetic field maps from MFM images involves a deconvolution requiring knowledge of the shape and magnetization configuration of the tip. MFM is possible under a wide variety of conditions, including in air, liquid, vacuum, and over a broad range of temperatures. Highest resolution and sensitivity, however, are achieved at cryogenic temperatures and in ultra-high vacuum.

*Sensitivity to different types of contrast*

Depending on the type of transducer and its tip, MFM maps magnetic field or magnetic field gradients. The ultimate noise limiting these measurements is thermal noise acting on the transducer. Such noise causes random fluctuations in the measured vibration amplitude and frequency. As shown in Box 1, thermal noise sets a minimum measurable magnetic field or field gradient, depending on the measurement type and the magnetization configuration of the tip. For example, a frequency shift measurement of a conventional MFM transducer [74] has a thermal limit at 4 K to static gradients of $\left(\frac{\partial B}{\partial r}\right)_{min} \approx 30$ T/(m Hz$^{1/2}$). Recently demonstrated NW MFM has a thermal limit for the same measurement that is about 1000 times smaller [16].

*Box 1: MFM*

Force microscopy contrast is generated by the interaction of a cantilever tip with the sample underneath. By monitoring the vibration amplitude, one can measure tip-sample forces at the cantilever resonance frequency, while by monitoring the vibration frequency, one can measure static tip-sample force gradients. The ultimate noise limiting these measurements is the thermal (Brownian) motion of the cantilever. Thermal noise sets a minimum measurable resonant force $F_{min} = \sqrt{4\,k_B T\,\Gamma}$ in an amplitude measurement and a minimum measurable static force gradient $\left(\frac{\partial F}{\partial r}\right)_{min} = \frac{1}{r_{rms}}\sqrt{4\,k_B T\,\Gamma}$ in a frequency measurement, where $k_B$ is the Boltzmann constant, $T$ is the temperature, $\Gamma$ is the mechanical dissipation, $r_{rms}$ is the cantilever oscillation amplitude, and $\hat{r}$ indicates the direction of cantilever oscillation.

In MFM, the magnetic tip transduces a magnetic field profile into a force profile. This interaction can often be approximated using a point-probe model, in which an effective magnetic multipole – including a monopole $q$ and a dipole $\boldsymbol{m}$ – represents the magnetization distribution of the tip. A magnetic field profile $\boldsymbol{B}$ then produces a magnetic force acting on the cantilever given by $\boldsymbol{F}_{MFM} = q\,\boldsymbol{B} \cdot \hat{r} + \nabla(\boldsymbol{m} \cdot \boldsymbol{B}) \cdot \hat{r}$. Note that, in most cases, the contribution of the torque generated by $\boldsymbol{B}$ is negligible. For conventional MFM, where the tip-sample interaction can be approximated by a pure magnetic monopole, this results in a minimum measurable resonant magnetic field $B_{min} = \frac{1}{q}\sqrt{4\,k_B T\,\Gamma}$ and a minimum measurable static magnetic field gradient $\left(\frac{\partial B}{\partial r}\right)_{min} = \frac{1}{q\,r_{rms}}\sqrt{4\,k_B T\,\Gamma}$. Purely dipolar tips, such as those on the ends of some NWs [15], are sensitive to a further spatial derivative of the magnetic field, compared to monopolar tips. Similar expressions can be written limiting those measurements.

By comparing the thermal noise background to the expected magnetic field or field gradient from a single Bohr magneton $\mu_B$ or a line or current $I$, as calculated in Box 2, we can assess the sensitivity of MFM. For example, conventional MFM scanning 50 nm above a sample is sensitive to frequency shifts equivalent to a magnetic moment of a few thousand $\mu_B$/Hz$^{1/2}$ or currents of a few µA/Hz$^{1/2}$ [74]. The same type of measurement carried out with newly demonstrated NW MFM probes 100 nm above a sample is about 100 times more sensitive to each type of contrast [15,16]. Estimates of sensitivity to magnetic moment and current as a function of probe-sample spacing are shown in Fig. 2.

It should be noted that the thermal limit on frequency measurements is rarely reached in practice. Most frequency measurements are limited by other noise sources, such as temperature variations, adsorption-desorption noise, or other microscopic mechanisms intrinsic to the resonator, that are typically an order of magnitude larger [75]. On the other hand, measurements of resonant oscillation amplitude, which are sensitive to modulations at the mechanical frequency of the sensor (typically in the 100 kHz regime), are often thermally limited. In such measurements, conventional MFM cantilevers can be sensitive to a few hundred $\mu_B$/Hz$^{1/2}$ or a few hundred nA/Hz$^{1/2}$, while NW MFM transducers reach down to a few $\mu_B$/Hz$^{1/2}$ or a few nA/Hz$^{1/2}$.

*Box 2: Magnetic field sources*

The magnetic field of a magnetic moment $\boldsymbol{m}$ at distance $\boldsymbol{r}$ is given by $\boldsymbol{B}_m = \frac{\mu_0}{4\pi r^3}\left(\frac{3\,(\boldsymbol{m}\cdot\boldsymbol{r})\boldsymbol{r}}{r^2} - \boldsymbol{m}\right)$ and the magnetic field of a line of current $\boldsymbol{I}$ is given by: $\boldsymbol{B}_I = \frac{\mu_0 \boldsymbol{I}\times\boldsymbol{r}}{2\pi r^2}$, where $\mu_0$ is the vacuum permeability. Using these two equations, we can express the various quantities measured by our scanning probe sensors as a function of tip-sample spacing in terms of $\mu_B$ of magnetic moment or A of current. For example, for SNVM measuring the z-component of the stray magnetic field, the maximum measurable signal from a single $\mu_B$ moment pointing along the z-direction at a tip-sample spacing $z$ is $B_{\mu_B,z} = \frac{\mu_0 \mu_B}{2\pi z^3}$, while the maximum from a line of current $I$ flowing in the plane is $B_{I,z} = \frac{\mu_0 I}{4\pi z}$. Similar expressions can be written for the maximum magnetic flux in the z-direction from the same moment and current measured by SSM: $\Phi_{\mu_B,z} = \frac{\mu_0 \mu_B R^2}{2(z^2+R^2)^{3/2}}$, where $R$ is the SQUID radius and $\Phi_{I,z} = \frac{\mu_0 I D}{4\pi} \ln\left(\frac{D^2+4z^2+D\sqrt{D^2+4z^2}}{D^2+4z^2-D\sqrt{D^2+4z^2}}\right)$, where $D$ is the length of one side of a square SQUID loop (to simplify the calculation, the current is integrated over a square rather than a circular loop). The corresponding maximum static magnetic field gradients measured by standard MFM are: $\frac{\partial B_{\mu_B,z}}{\partial z} = \frac{3\mu_0 \mu_B}{2\pi z^4}$ and $\frac{\partial B_{I,z}}{\partial z} = \frac{3\sqrt{3}\mu_0 I}{16\pi z^2}$.

*Applications to 2D materials*

High-sensitivity MFM probes could visualize correlated states in 2D systems via frequency shift maps, which can ultimately be reconstructed in current density or magnetization contrast. This would be particularly interesting for measuring the spatial localization of flowing currents, as in edge states, and for the determination of length scales such as magnetic domain sizes and coherence lengths. Visualizing current flow in MATBG [6] and WeT$_2$ [76,77] while they are electrostatically tuned into their superconducting states, would help reveal the origin of this superconductivity and whether or not it is topological. NW MFM may also help provide direct evidence for magnetism in 2D magnets or even in the 2D semiconductor, monolayer MoS$_2$ [78,79]. Optical spectroscopy has provided evidence of a high-field spin-polarized state in this material, however, confirmation of its presence via a direct measurement of magnetic field has not yet been possible. NW MFM's high sensitivity and ability to operate in high-field conditions make it promising for such an investigation.

MFM can also be used to map dissipation in a sample by measuring the power required to maintain a constant oscillation amplitude. This type of contrast maps the energy transfer between the tip and the sample and provides excellent contrast for nanometer-scale magnetic structure [80]. Since energy dissipation plays a central role in the breakdown of topological protection, it may provide important contrast in spatial studies of strongly correlated states in 2D vdW materials. Dissipation contrast has been used to observe superconducting [81] and bulk structural phase transitions [82], as well as the local density of states. 2D materials engineering allows for the fabrication of devices, in which a variety of different physical phases can be accessed by the application of a gate voltage. Local measurements of dissipation via MFM could be an important tool for making spatial maps of the transitions between those states.

## Scanning SQUID microscopy

### Working principle and conditions

In scanning SQUID microscopy (SSM), contrast results from the magnetic flux threading through a superconducting loop that is interrupted by at least one JJ. The SQUID's critical current is periodic in this flux – given by the magnetic field integrated over the area of the loop $\Phi_z = \int \boldsymbol{B} \cdot \boldsymbol{dA}$ – with a period given by the flux quantum $\Phi_0$. By applying the appropriate current bias, one can detect voltages across the SQUID which correspond to changes in magnetic field threading the SQUID loop corresponding to factions of a $\Phi_0$, typically down to $10^{-6}$ $\Phi_0$/Hz$^{1/2}$. For imaging applications, a DC SQUID with two JJs is most often used. This loop – or a pick-up loop inductively coupled to it – is scanned above a target sample in order to map the magnetic field profile. The loop's size is minimized in order to optimize spatial resolution. SQUIDs operate only below a superconducting transition temperature, which is typically below 10 K, but can be above the temperature of liquid nitrogen (77 K) for some high-$T_c$ superconductors.

### Sensitivity to different types of contrast

The noise limiting the measurement of magnetic flux in a SQUID arises from several sources including Johnson noise, shot noise, 1/f noise, and quantum noise [73]. For SQUIDs smaller than 1 µm and at frequencies high enough to avoid 1/f noise, quantum noise sets the fundamental limit on detectable flux to be $\Phi_Q = (\hbar L)^{1/2}$, where $\hbar$ is Planck's constant and $L$ is the loop inductance [73,83,84]. State-of-the-art SOT sensors made from Pb combine the highest flux sensitivity with the smallest sensor size. In the white-noise limit (measured in the kHz range), sensors with 50 nm diameter reach $\Phi_{min} = 50$ n$\Phi_0$/Hz$^{1/2}$, which is about 4 times larger than $\Phi_Q$ [32]. Near DC (measured in the Hz range), where the same sensor is limited by 1/f noise, $\Phi_{min}$ is about 10 times larger. In these devices, $L$ is dominated by kinetic rather than geometric inductance. For this reason, optimizing material parameters for low kinetic inductance provides the best route for improving $\Phi_{min}$.

What this sensitivity means in terms of magnetization or current sources requires knowing the tip-sample spacing. Using the best 50-nm-diameter SOT at a spacing of 50 nm – closer approach than the characteristic sensor size does not improve spatial resolution – the white noise level is equivalent to the field of a few $\mu_B$/Hz$^{1/2}$ or a few tens of nA/Hz$^{1/2}$, while at DC the device is ten times less sensitive. Again, such estimates are shown as a function of probe-sample spacing in Fig. 2.

### Applications to 2D materials

SSM has already been successfully used to image current density via local measurements of Biot-Savart fields. In particular, maps of the flow of equilibrium currents in graphene made using SOT probes revealed the topological and non-topological components of edge currents in the quantum Hall state [9]. In MATBG, similar scanning probe measurements revealed the degree of twist-angle disorder [8] and magnetization images have provided evidence for orbital magnetism [11]. Given the SOT's exquisite sensitivity to local temperature, these probes can also be applied to measure local sources of dissipation, as was demonstrated in experiments on graphene [40,85]. Similar scanning probe measurements of magnetic field and dissipation could be carried out on other moiré systems, including twisted transition metal dichalcogenides and twisted multi-layer graphene. These systems are also predicted to host a variety of correlated states, including superconductivity, Mott insulating states, magnetic states, and Wigner crystal states [86].

## Scanning NV center microscopy

### Working principle and conditions

In scanning NV-center microscopy (SNVM), magnetic field measurements are carried out by using optically-detected magnetic resonance (ODMR) spectroscopy, where the EPR spectrum of the NV is recorded by simultaneous microwave excitation and optical readout of the defect's spin state as the probe is scanned in close proximity to the sample surface. Thanks to the technique of single-molecule fluorescence, these experiments can be performed on a single spin [87]. The magnetic field sensitivity results from a Zeeman shift of the spin resonances. In the regime of a weak orthogonal component of an external magnetic field, the field component parallel to the NV symmetry axis leads to a linear shift of the $m_s = \pm 1$ spin states with a proportionality given by the free-electron gyromagnetic ratio $\gamma = 2\pi \times 28$ GHz/T [88]. The ODMR spectrum is measured as a change in optical intensity as a function of continuous-wave or pulsed microwave excitation [89]. Other forms of contrast include ODMR quenching in magnetic fields larger than 10 mT due to energy-level mixing by the off-axis field component [71,72] and spin relaxometry [83]. The latter probes high-frequency fluctuations near the NV resonance (GHz range) and allows for the investigation of magnetic fluctuations and spin waves in ferromagnets [90–92]. Further, dynamical decoupling techniques can be used to perform frequency spectroscopy in the kHz-MHz range [93,94].

In scanning probe applications, the NV center is hosted within a crystalline diamond nanopillar and scanned over the sample of interest [41,95]. State-of-the-art diamond probes are engineered with shallow NV centers, which are implanted at depths around 10 nm [96], in order to minimize the distance between the NV center and the sample and thus to optimize both sensitivity and spatial resolution. However, in most SNVM literature, the NV stand-off distance is 50 to 100 nm, indicating that NV centers may be deeper than expected.

### Sensitivity to different types of contrast

SNVM is typically limited by photon shot noise from the optical readout, and can be expressed by a simple signal-to-noise formula typical for optical magnetometry [97]. Specifically, the magnetic sensitivity of the scanning NV magnetometer is determined by a combination of the spin dephasing or decoherence time $T_2$, the optical contrast $\epsilon$ and the maximum photon count rate $I_0$. A generic estimate for the minimum detectable magnetic field is given by $B_{\min} \approx \left[\gamma\epsilon\sqrt{I_0 t_{acq} T_2}\right]^{-1}$, where $\gamma$ is the gyromagnetic ratio and $t_{acq}$ is the photon integration time. Using typical values ($\epsilon = 0.2$, $I_0 = 200$ kC/s, $t_{acq} = 300$ ns, $T_2 = T_2^* = 1.5$ µs), the minimum detectable field is about 1 µT/Hz$^{1/2}$ for pulsed operation and 10 µT/Hz$^{1/2}$ for continuous-wave operation. Recent SNVM experiments have shown state-of-the-art pulsed sensitivity of 100 nT/Hz$^{1/2}$ [45]. In the future, the sensitivity can be improved by extending $T_2$ using isotopically-purified (free of $^{13}$C) material [46] and AC magnetometry techniques [98], improving the contrast through alternative readout schemes [47], and improving the count rate by photonic shaping [99,100].

If we assume the best demonstrated pulsed sensitivity and a 25 nm NV-sample distance, SNVM is sensitive to one $\mu_B$/Hz$^{1/2}$ or a few tens of nA/Hz$^{1/2}$. Fig. 2 shows such sensitivity estimates for some of the best SNVM as a function of probe-sample spacing.

### Applications to 2D materials

SNVM has been applied to image magnetization in the 2D ferromagnets [12–14] and current flow in graphene [61,62,101] and layered semimetals [45]. Given SNVM's particularly high sensitivity to magnetic moment, the technique is particularly suited for mapping magnetism in vdW magnets to distinguish domain structure, quantify the strength of the magnetism, and confirm its origin. The ability to distinguish the magnetism of single atomic layers, as first shown in CrI$_3$ [12] and later in CrBr$_3$ [13] and CrTe$_2$ [14], is crucial for investigating the effect of each layer in vdW heterostructures. The ability of SNVM to retain high sensitivity at room temperature and under ambient conditions makes it applicable to magnetic systems with potential practical application in spintronic devices. High-frequency sensing with SNVM [102] may also be useful for investigating magnonic excitations in 2D magnets.

Although current mapping at temperatures below 4 K, such as required for studies of superconductivity in 2D materials, is still challenging, SNVM is ideal for experiments across a broad and higher temperature range. In fact, researchers have used SNVM to map hydrodynamic flow in graphene [62] and WTe$_2$ [45], which is strongest at intermediate temperatures. Similar studies could be carried out in a plethora of other 2D systems, in which viscous electron transport may dominate under certain conditions.

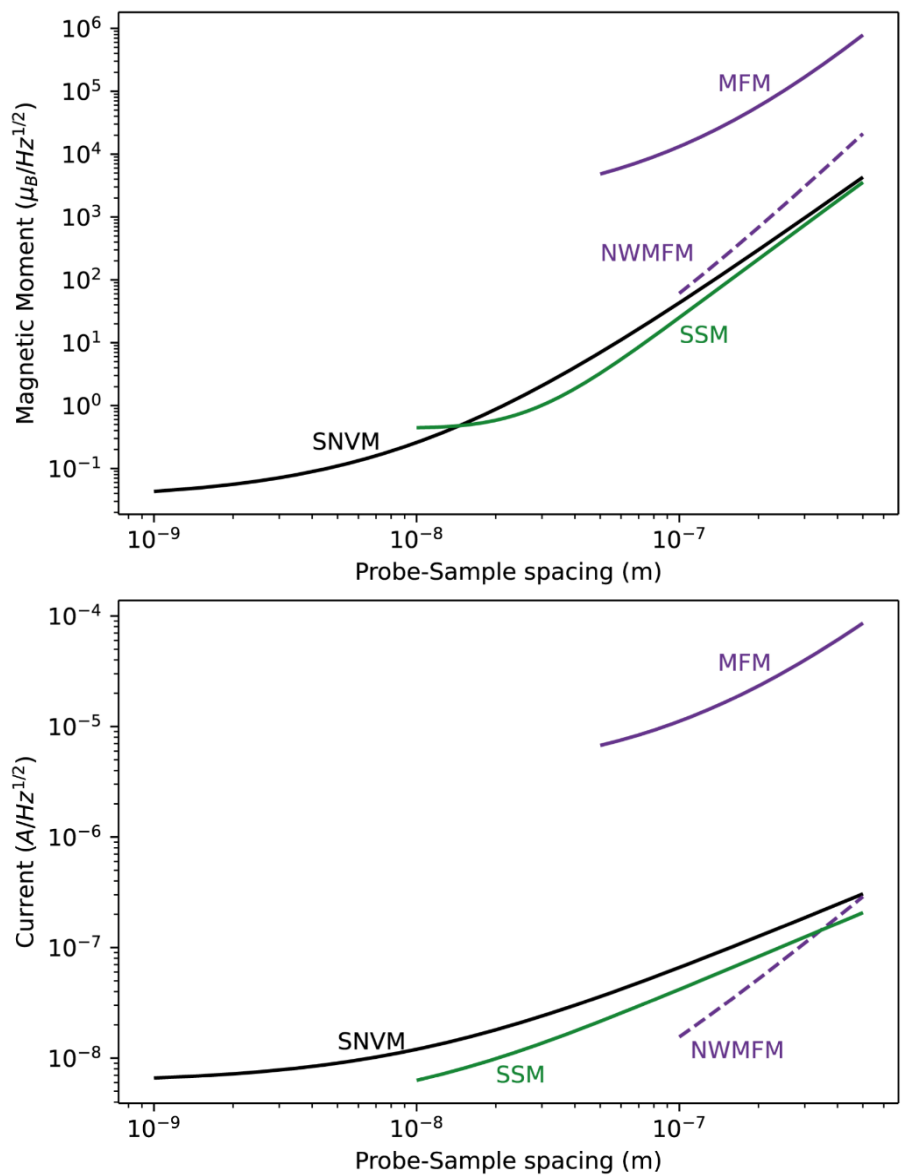

Figure 2: Plots comparing the sensitivity to magnetic moment (top) and current (bottom) of the 3 magnetic imaging techniques under the most favorable conditions, i.e. in vacuum and at liquid helium temperatures. We use parameters from van Schendel et al. for conventional MFM [74], Mattiat et al. for NW MFM [16], Vasyukov et al. for SSM [32], and Vool et al. for SNVM [45]. MFM and NW MFM

sensitivities are based on frequency shift measurements at DC, while SNVM and SSM sensitivities are based on AC measurements usually in the tens of kHz range.

## Comparison between techniques

Having quantified the sensitivity of MFM, SSM, and SNVM to magnetic moment and electrical current, we can now see which techniques are best suited for mapping which type of contrast. Fig. 2 shows the sensitivity of all techniques to the magnetic field profile produced by a magnetic moment and a line of current as a function of probe-sample spacing. In the case of conventional and NW MFM, we refer to thermal limit of frequency shift measurements, which applies to DC or low-frequency measurements. In the other two cases, we use the minimum flux and field noise achieved in these devices in AC measurements in the tens of kHz range.

|  | MFM (conventional) [19,20,74,103] | MFM (NW) [16] | SSM (susceptometer) [29] | SSM (SOT) [32] | SNVM [43–45,65] |
|---|---|---|---|---|---|
| Sensor size | 10-100 nm | 100 nm | 0.5 µm | 50 nm | < 1 nm |
| Sensor stand-off | 10-100 nm | 50 nm | 330 nm | 25 nm | 50 nm |
| Spatial resolution | 10-100 nm | 100 nm | 0.5 µm | 100 nm | 15-25 nm |
| DC sensitivity | 10-100 µT/(Hz)$^{1/2}$ | 3 nT/(Hz)$^{1/2}$ | 660 nT/(Hz)$^{1/2}$ | 50 nT/(Hz)$^{1/2}$ | 4 µT/(Hz)$^{1/2}$ |
| AC sensitivity | 170 nT/(Hz)$^{1/2}$ | 3 nT/(Hz)$^{1/2}$ | 130 nT/(Hz)$^{1/2}$ | 5 nT/(Hz)$^{1/2}$ | 100 nT/(Hz)$^{1/2}$ |
| Operating field | < 10 T | < 10 T | < 100 mT | < 1.2 T | < 100s mT |
| Operating temp. | < 500 K | < 300 K | < 9 K | < 7 K | < 600 K |

Table 1: Parameters for state-of-the-art magnetic SPM combining the highest-sensitivity with the highest resolution, based on the devices discussed in the cited references. Values shown in gray represent estimates based on the properties of the sensors, which have not yet been experimentally confirmed.

Together with sensor size, probe-sample spacing sets the spatial resolution of an SPM technique. Depending on the type of contrast, this spacing also strongly affects sensitivity. SSM sensitivity is not shown closer than 10 nm, because sensors are difficult to operate closer without a catastrophic crash. MFM sensitivity is not shown closer than 50 nm and NW MFM is not shown closer than 100 nm, because the point-probe approximation breaks down at tip-sample spacings smaller than the tip size and non-contact friction starts to dominate the force noise [104]. Also, at such close spacing, the stray field produced by the MFM tip at the sample is often invasive. Since SNVM can essentially be operated in contact with the sample, we plot its sensitivity down to 1 nm of probe-sample spacing.

Depending on tip-sample spacing, either SNVM or SSM have the highest sensitivity to magnetic moment. SSM appears best for tip-sample distances larger than 25 nm, while SNVM is better for closer approach. Conventional MFM is the least sensitive, while NW MFM is competitive with the other techniques. While very promising, NW MFM tip size must be reduced from state-of-the-art diameters of 100 nm in order for the technique to become competitive in high spatial resolution imaging of magnetic moment.

Among proven techniques, SSM is most sensitive to current. While conventional MFM is the least sensitive, NW MFM appears to surpass all techniques between 500 and 50 nm. Once again, for spatial resolutions better than 10 nm SNVM appears to be the best choice.

Fig. 3 provides another way to compare the three techniques, by showing the characteristic length of each sensor (its size in one dimension) together with its sensitivity to magnetic field. We plot a few state-of-the-art sensors of each type and give an approximate idea of each technique's operating regime. The characteristic length of a sensor not only sets its ultimate spatial resolution, but also sets the optimum probe-sample spacing, since closer approach is either impossible or does not improve sensitivity. Diagonal lines represent the combined probe-sample spacing and field noise required to achieve a certain sensitivity to magnetic moment or current.

Fig. 3 makes clear that SNVM has the smallest characteristic length, due to the atomic-scale of the NV center and the possibility to implant NVs with long coherence times just 10 nm from the surface of a scanning probe. This makes SNVM the technique of choice for spatial resolution under 25 nm and for the detection of small magnetizations. Because the magnetic field produced by a magnetic moment drops of with the inverse cube of the probe-sample distance, a small sensor able to work in close proximity to the sample is crucial for this type of contrast.

Fig. 3 also shows that SSM has the highest field sensitivity, but that it comes at the expense of large sensor size. While conventional MFM appears too insensitive to measure weak magnetization or current density, the increased force sensitivity of NW MFM makes it competitive with the other two techniques. In fact, for the measurement of currents, where spatial resolutions better than 100 nm are not required, SSM and NW MFM are the best techniques. Because Biot-Savart fields fall off only with the inverse power of the probe-sample spacing, a small sensor is not as important in current measurements as it is in magnetization measurements.

Aside from their sensitivity and resolution, each technique has properties making it more or less advantageous for certain samples. The strongly magnetic tip of an MFM can produces tens of mT of magnetic field on a sample 50 nm away. This field can in turn perturb the sample, potentially altering its state. SNVM requires the excitation of the probe with visible laser light. This optical excitation can perturb optically active samples below the probe. On the other hand, the stray fields due to the Meissner effect on an SSM probe are nearly negligible, making these sensors minimally invasive. SSM, however, is the most limited from the environmental point of view, functioning only at temperatures below the superconducting transition of the SQUID, typically below 10 K. Both MFM and SNVM function at a wide range of temperatures and pressures. SSM must also work below its critical field, which for state-of-the-art SOTs can be as high as a few T. SNVM is also limited in field, in that the frequency of the microwaves used to address the NV center scale linearly with field and become impractically high above 1 T.

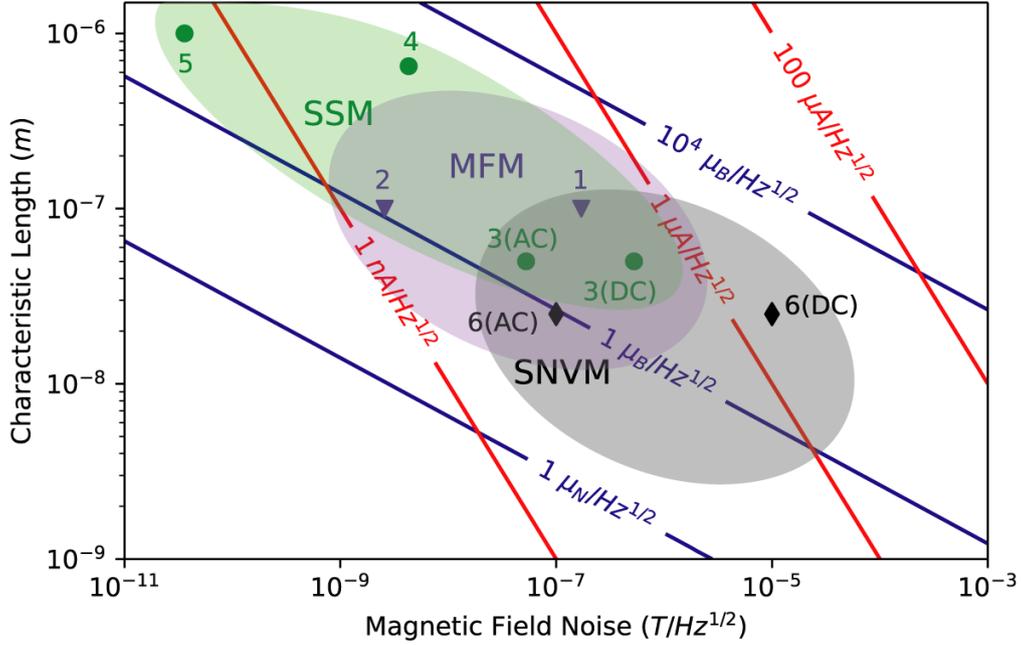

Figure 3: Plot showing the characteristic length and magnetic field noise of state-of-the-art scanning magnetic probes under ideal conditions, i.e. in vacuum and at liquid helium temperatures. The characteristic length sets the scale of the possible spatial resolution. Diagonal lines show the sensitivity required to measure the labelled magnetic moments and currents. Data points correspond to state-of-the-art SPMs demonstrated in the corresponding reference: 1, van Schendel et al. [74]; 2, Mattiat et al. [16]; 3, Vasyukov et al. [32]; 4, Kitley et al. [73]; 5, Jeffery et al. [105]; 6, Vool et al. [45].

### Reconstruction of magnetization or current from field images

Since magnetic field microscopy techniques do not directly image the current or magnetization pattern, but rather their stray field, the question arises whether and how the former may be reconstructed from a stray field map. The relation between stray field and current density is governed by the Biot-Savart law that, via the concept of bound currents, can also be applied to magnetization.

Work in the late eighties by Roth [106] and Beardsley [107] established a framework to compute the stray fields of two-dimensional current density $J(x,y)$ and two-dimensional magnetization patterns $M(x,y)$, respectively. The same work also specified the conditions, in which a reconstruction of $J$ and $M$ is possible. In particular, they showed that three-dimensional current densities and magnetization patterns do not produce a unique magnetic stray field pattern, and can therefore not be determined by stray field imaging. Further, even an arbitrary two-dimensional magnetization pattern does not possess a unique stray field because the divergence-free part of $M$ does not generate an external stray field and is left arbitrary [107]. A rigorous solution, on the other hand, exists for two-dimensional current densities $J = (J_x, J_y, 0)$ and out-of-plane magnetized films $M = (0, 0, M_z)$. It has further been shown that this solution can be extended to thick films if the magnetization, or current density, is uniform through the thickness [43]. As a consequence, magnetic field imaging is especially useful for analyzing 2D systems and thin-film devices.

Magnetic field maps do not reproduce all current or magnetization features with the same sensitivity. Looking at the mechanics of the reconstruction, shown in Box 3, it becomes clear that features smaller than the probe-sample spacing $z$ produce negligible magnetic field at the sensor location, because stray fields decay exponentially with distance from the surface. The decay length is given by $\lambda/2\pi$, where $\lambda$ is the spatial wavelength of the current or magnetization feature, as shown in Fig. 4. Interestingly, large features compared to the probe-sample spacing, i.e. large $\lambda/z$, produce a strong signal for currents, but not for magnetization.

Imaging magnetic field gradients rather than magnetic fields, allows one to push the maximum sensitivity towards smaller feature size. Magnetic gradient detection is the standard mode for MFM, but can also be implemented for SSM and SNVM by a mechanical oscillation of the sensor [9,41]. Using lock-in techniques to demodulate the resulting signal can also significantly reduce noise through spectral filtering. Gradient detection is especially attractive for imaging currents, because the magnetic gradient image closely resembles the current density image, so that no reconstruction is needed [9]. For SNVM, gradient imaging is attractive because it upconverts DC signals to AC where much more sensitive magnetometry protocols are available [41,98].

*Box 3: Reconstruction of current density and magnetization from a magnetic field image*

The current density $\boldsymbol{J} = (J_x, J_y)$ and in-plane magnetization $M_z$ of a two-dimensional sample can be conveniently reconstructed from a magnetic field image by expressing the Biot-Savart law in $k$-space. Assume that we image in a plane at distance $z$ above the sample, the magnetic stray field, in $k$-space is given by: $B_z(k_x, k_y, z) = ig(k,z)[\frac{k_y}{k}J_x(k_x, k_y) - \frac{k_x}{k}J_y(k_x, k_y)]$, where $g(k,z) = \frac{1}{2}\mu_0 d e^{-kz}$ is a transfer function with $d \ll z$ being the film thickness, $k_x$ and $k_y$ are the $k$-vectors, and $k = (k_x^2 + k_y^2)^{1/2}$. Similar expressions can be derived for $B_x$ and $B_y$ as well as for $d \geq z$ [43,106]. To reconstruct the current density from a magnetic field map, the relation is inverted: $J_x(k_x, k_y) = -\frac{ik_y W B_z(k_x,k_y,z)}{kg(k,z)}$ and $J_y(k_x, k_y) = -\frac{ik_x W B_z(k_x,k_y,z)}{kg(k,z)}$, where $W$ is a window function, whose cut-off wavelength is adjusted to suppress high-frequency noise. Different choices for the window function have been reported in the literature, including Hann and rectangular and Tikhonov-based windows. The cut-off wavelength typically is of order $z$. An expression for reconstructing $\boldsymbol{J}$ from an arbitrary $B$-field component is given in [43].

Similar expressions can be derived for reconstructing an out-of-plane magnetization $M_z$ or to reconstruct magnetic gradient images. To reconstruct $M_z$, note that $\boldsymbol{J} = \nabla \times \boldsymbol{M}$, and therefore: $B_z(k_x, k_y) = kg(k,z)M_z(k_x, k_y)$ for the forward problem as well as $M_z(k_x, k_y) = \frac{W B_z(k_x,k_y,z)}{kg(z,k)}$ for the reverse problem. To reconstruct a magnetic gradient image, the transfer function incurs an additional factor of $k$ due to the derivative.

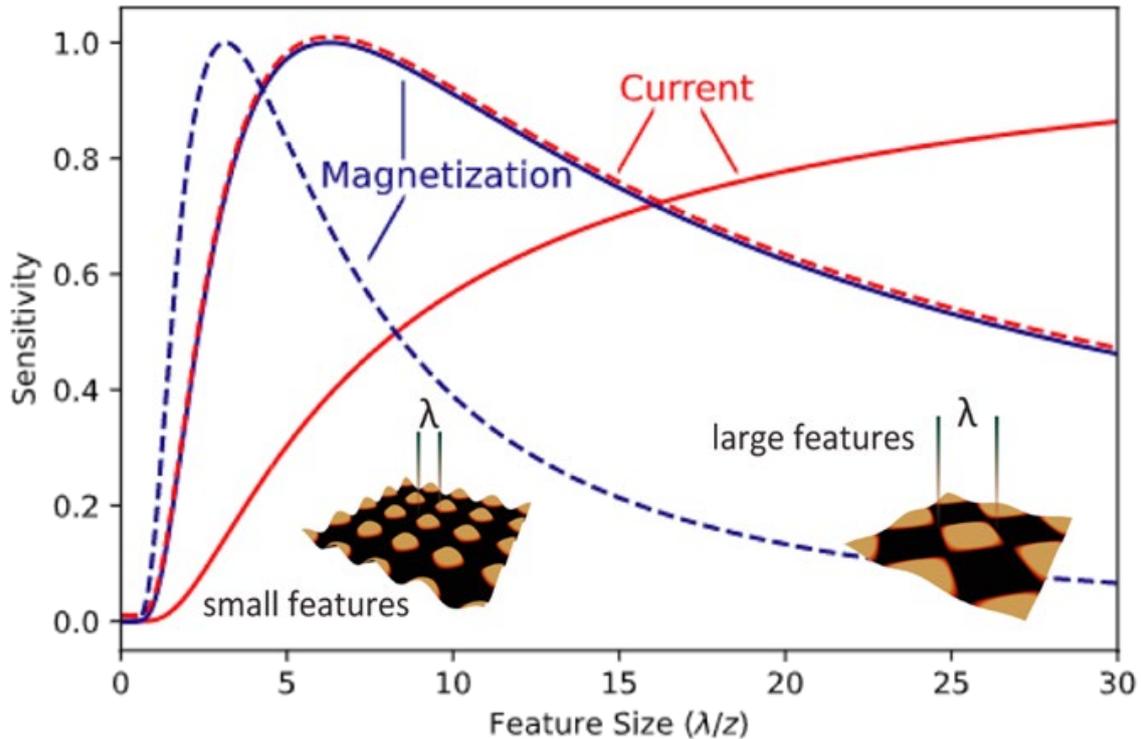

Figure 4: Sensitivity as a function of feature size, expressed as the ratio between the feature's spatial wavelength $\lambda$ and the probe-sample spacing $z$. (Solid lines) Magnetic field imaging is most sensitive to spatially large current features (red) and to magnetization features (blue) with a size similar to the probe-sample spacing $z$. (Dashed lines) Magnetic gradient imaging shifts the maximum sensitivity towards smaller feature size.

## Prospects for improvement

Improving MFM sensitivity requires stronger magnetic tips or transducers with better force sensitivity. Up to an order of magnitude in force sensitivity could be gained by using optimized NW transducers. MFM cantilevers have recently been realized with spring constants in the hundreds of mN/m and mechanical quality factors above $10^6$, resulting in nearly 100 times more sensitivity than conventional transducers. In general, however, improving the sensitivity of a mechanical transducer is achieved by reducing its size [108], as in recent work on NW MFM. Another route to improve magnetic field sensitivity is to increase the magnetic moment and size of MFM tips. This gain, however, comes at the cost of reducing spatial resolution and increasing the perturbative effect of the probes, which now produce larger stray fields at the sample.

The spatial resolution of the MFM could be improved by utilizing the sharpest possible magnetic tips. Extensive work has been done in this area in the context of conventional MFM, achieving spatial resolutions down to 10 nm [109–112]. Such work could be extended to high-force-sensitivity NW MFM. Smaller tips, however, have reduced magnetic moment and, consequently, a worse sensitivity to magnetic field profiles. In order to maintain high sensitivity, in general, the reduction in tip size should be accompanied with a reduction in transducer size.

Improvements in SSM field sensitivity could come from a reduction in the SQUID inductance. Given that this quantity is dominated by kinetic inductance in state-of-the-art devices, optimizing the superconducting material from which the device is made could be a fruitful pursuit. Further reduction of the characteristic size of SSM probes is difficult to imagine. SOT probes have been fabricated with diameters just under 50 nm. Reducing this size further would make the device size similar to the thickness of the deposited superconducting film, complicating much of the process, on which the fabrication is based. SQUIDs with feature sizes of only a few nanometers have been fabricated in YBCO using a focused ion beam of He [113], raising the possibility of devices that are an order of magnitude smaller and potentially work at liquid nitrogen temperature. Nevertheless, significant work remains to be done before such devices can be integrated onto scanning probes.

In order to reduce the characteristic length scale of SNVM, a number of researchers have focused on simultaneously reducing the implantation depth of NV centers and maintaining their coherence properties. Implantation depths of less than 3 nm have been reported combined with greater than 10 µs coherence times [114], giving a perspective of better than 10 nm imaging resolution combined with sub-10 nT/Hz$^{1/2}$ sensitivity. So far, however, most reported stand-off distances remain between 50 and 100 nm and the best magnetic field sensitivities at 100 nT/Hz$^{1/2}$ and significant work may be needed to reduce either figure of merit.

## Conclusion

The confluence of substantial improvements in nanometer-scale magnetic imaging with the advent of engineered 2D materials creates the perfect opportunity to gain new insight into the physics of correlated states in condensed matter. The unprecedented control provided by layer-by-layer material engineering gives physicists a vast playground on which to test theories on superconductivity, magnetism, and other correlated phenomena. With this control, however, comes sensitivity to disorder and inhomogeneity. In such a fragile environment, local measurements – with sensors whose characteristic size is smaller than the length scale of the disorder – are essential for making sense of the system. For this reason, SPM techniques will become ever more important tools in this growing field, perhaps only losing traction, once fabrication techniques have been honed and substantially improved.

There are a number of SPM techniques, which have emerged as important tools for the investigation of 2D systems. Conventional atomic force microscopy has been used extensively for topographic characterization of 2D materials. In graphene, scanning single electron transistors have been used to map the local density of states [115] and for imaging hydrodynamic flow [116]. Scanning gate microscopy has been used to image localized states [117] and scanning microwave impedance microscopy for visualizing the structural details of moiré lattices [118]. Electronic properties of 2D transition metal dichalcogenides have also been studied by scanning tunnelling microscopy [119]. Scanning near-field optical microscopy has even been used to measure polaritonic response in graphene-hexagonal boron nitride heterostructures [120].

As discussed in this review, among these SPM techniques, those involving non-invasive magnetic field imaging are particularly suited to investigating the correlated states present in 2D systems, because of their ability to map both current and out-of-plane magnetization. Given the high sensitivity and spatial resolution required to investigate correlated states in 2D materials, it is important to choose the appropriate magnetic SPM for the physical system under investigation. The different scaling of magnetization and current contrast with probe-sample spacing and the different physical quantities that are measured by various magnetic SPM make certain techniques more amenable to certain systems. We hope to have provided some insight in this regard, both to experimentalists wanting to apply magnetic SPM to 2D systems and to physicists working on the next generation of magnetic imaging techniques.

## Acknowledgements

We thank Prof. Hans-Josef Hug for insightful discussions. We acknowledge the support of the Canton Aargau and the Swiss National Science Foundation under Project Grant 200020-178863, via the Sinergia Grant "Nanoskyrmionics" (Grant No. CRSII5-171003), and via the NCCR "Quantum Science and Technology" (QSIT). C.L.D. acknowledges funding by the Swiss National Science Foundation under Project Grant 20020-175600, by the European Commission under grant no. 820394 "ASTERIQS", and by the European Research Council under grant 817720 "IMAGINE".